\documentclass{aa}
\usepackage[varg]{txfonts}
\usepackage{graphicx}
\usepackage{natbib}
\bibpunct{(}{)}{;}{a}{}{,} 
\usepackage{color}

\begin{document}

\title{Solar activity during the Holocene: the Hallstatt cycle and its consequence for grand minima and maxima}

\author{I.G. Usoskin\inst{1}
\and Y. Gallet\inst{2}
\and F. Lopes\inst{2}
\and G. A. Kovaltsov\inst{3,4}
\and G. Hulot\inst{2}
}

\institute{Sodankyl\"a Geophysical Observatory (Oulu unit) and ReSoLVE Center of Excellence, University of Oulu, Finland
\and Institut de Physique du Globe de Paris, Sorbonne Paris Cit\'e, Universit\'e Paris Diderot, UMR 7154 CNRS, F-75005 Paris, France
\and Ioffe Physical-Technical Institute, 194021 St. Petersburg, Russia
\and Institute of Physics of the Earth, Russian Academy of Sciences, Moscow, Russia
}

\date{}

\abstract {}
{Cosmogenic isotopes provide the only quantitative proxy for analyzing the long-term solar variability
  over a centennial timescale.
While essential progress has been achieved in both measurements and modeling of the cosmogenic proxy,
 uncertainties still remain in the determination of the geomagnetic dipole moment evolution.
Here we aim at improving the reconstruction of solar activity over the past nine millennia using
 a multi-proxy approach.
}
{We used records of the $^{14}$C and $^{10}$Be cosmogenic isotopes, current numerical models of
 the isotope production and transport in Earth's atmosphere, and available geomagnetic field reconstructions,
  including a new reconstruction relying on an updated archeo-
and paleointensity database.
The obtained series were analyzed using the singular spectrum analysis (SSA) method to study the millennial-scale trends.}
{A new reconstruction of the geomagnetic dipole field moment, referred to as GMAG.9k, is built for the last nine millennia.
New reconstructions of solar activity covering the last nine millennia, quantified in terms of sunspot numbers, are presented and analyzed.
A conservative list of grand minima and maxima is also provided.}
{The primary components of the reconstructed solar activity, as determined using
 the SSA method, are different for the series that are based on $^{14}$C and $^{10}$Be. This shows that these primary components can only be ascribed
 to long-term changes in the terrestrial system and not to the Sun.
 These components have therefore been removed from the reconstructed series.
In contrast, the secondary SSA components of the reconstructed solar activity are found to be dominated by a common $\approx 2400$-year quasi-periodicity,
 the so-called Hallstatt cycle,  in both the $^{14}$C and $^{10}$Be based series. This Hallstatt cycle thus appears to be related to solar activity.
Finally, we show that the grand minima and maxima occurred intermittently over the studied period,
 with clustering near highs and lows of the Hallstatt cycle, respectively.
}

\maketitle

\section{Introduction}

Solar activity, which generically includes different manifestations of fluctuating processes
 in the solar convection zone on the surface and in the corona, varies on timescales from seconds to millennia.
Its long-term variability, with timescales longer than several centuries,
 can only be studied using indirect proxies, such as cosmogenic radionuclides \citep[e.g.,][]{beer12,usoskin_LR_13}.
For this purpose, the most useful cosmogenic isotopes are radiocarbon $^{14}$C and beryllium $^{10}$Be.
These isotopes allow reconstructing the past solar activity over the Holocene period, which spans the past 11 millennia, that
is,  since the last deglaciation.
Recently, several long-term solar activity reconstructions have been published
 \citep[e.g.,][]{solanki_Nat_04,vonmoos06,muscheler07,usoskin_AA_07,steinhilber12}.
They revealed variability in the solar activity on centennial and millennial scales, ranging from
 very low (almost spotless Sun) to high activity levels.
However, since these reconstructions rely on slightly different basic models and different datasets,
 they sometimes differ in details and overall levels.
In particular, although the existence of grand minima and maxima in solar activity has been known for a long
 time \citep[see the review of][]{usoskin_LR_13},
 it has remained a matter of debate whether grand minima and maxima are separate activity modes of the solar
 dynamo or simply non-Gaussian tails of its variability \citep[e.g.,][]{moss08,passos14a,karak15}.

To settle this question, a new approach was recently developed \citep{usoskin_AAL_14}, which takes into
 account the full range of uncertainties associated with a modern reconstruction of the
 $^{14}$C global production rate \citep{roth13}, an accurate millennial-scale
 archeomagnetic field reconstruction \citep{licht13}, and a detailed $^{14}$C production model
 \citep{kovaltsov12}.
Based on an analysis of the probability density function, the resulting reconstruction,
  limited to the past 3000 years \citep{licht13}, made it possible for the first
 time to show that grand minima of solar activity correspond to a distinct operational mode of the solar
 activity \citep{usoskin_AAL_14}.
Because of poor statistics, however, the result was inconclusive for the nature of the grand maxima.
Constraining the solar activity over longer period is more problematic, in particular because of uncertainties
 acknowledged before 500-1000 BC in paleo- and archeomagnetic
 field reconstructions \citep{snowball07}.
However, a large set of new archeo- and paleointensity data was acquired in the past few years
 (see Appendix A),
 which, as we argue, can improve our knowledge of the geomagnetic dipole moment evolution over most of the Holocene.
Here we take advantage of the new data to better constrain the long-term solar activity, as revealed
 from the use of a synthetic index of relative sunspot numbers.
We first extend the approach of \citet{usoskin_AAL_14} to the last 9 millennia using the reconstruction of the
 $^{14}$C global production rate \citep{roth13}, the
 $^{10}$Be GRIP dataset \citep{yiou97}, a new dipole moment reconstruction hereafter
 referred to as GMAG.9k (see Appendix A), and $^{14}$C and $^{10}$Be production models \citep{kovaltsov12,kovaltsov10}.
The recovered reconstructions aim at scrutinizing the variability in the solar activity over
 millennial and centennial timescales.
We also discuss the robustness of our new solar activity reconstructions using the results
 derived from other recent archeo- and paleomagnetic Holocene field models.
Finally, we present new results providing important observational constraints on the solar dynamo.

\section{Data}
\label{Sec:data}

\subsection{Cosmogenic radionuclide records}
\label{S:14C}
\begin{figure}[t]
\centering \resizebox{\columnwidth}{!}{\includegraphics{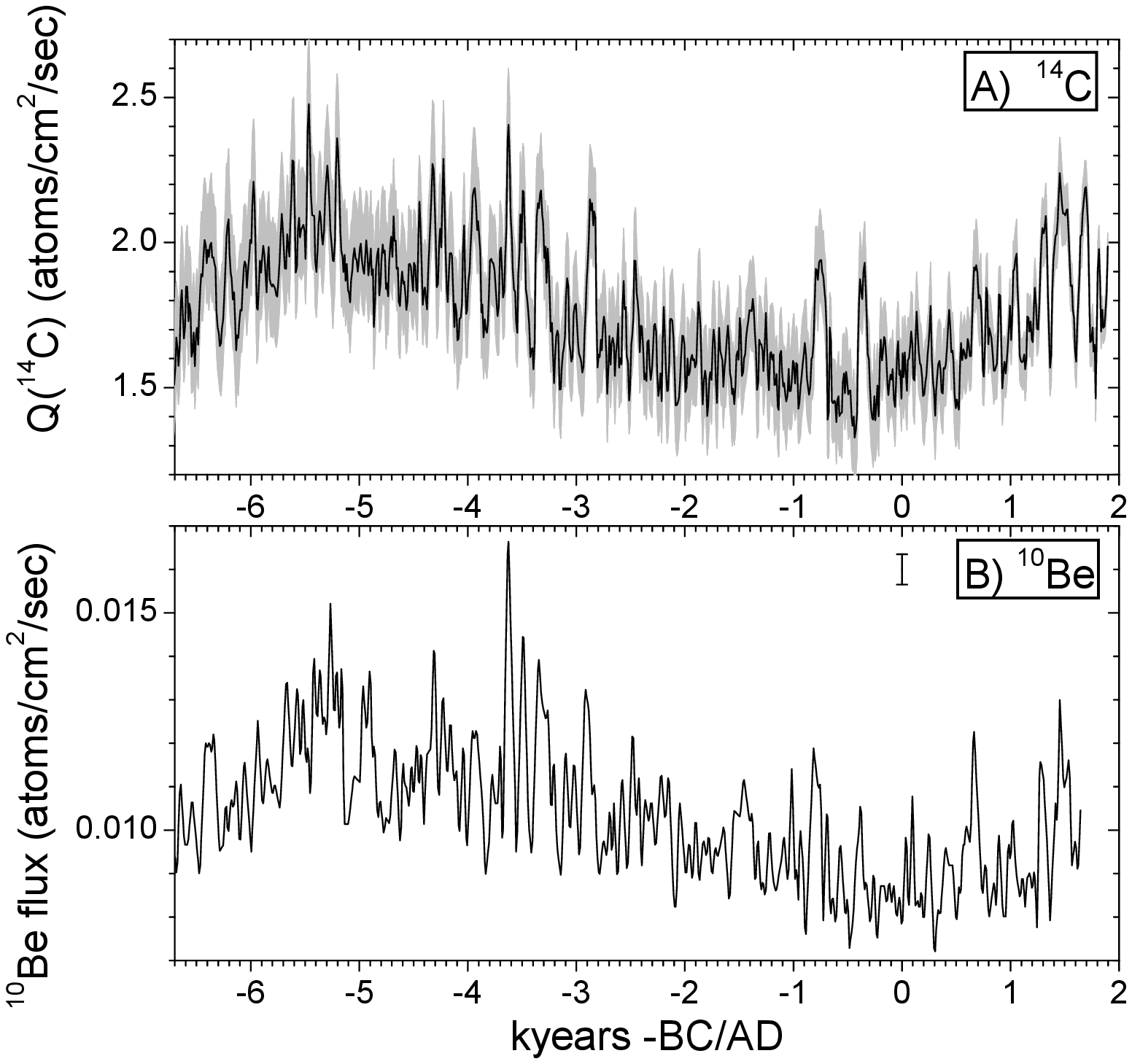}}
\caption{Time series of cosmogenic radionuclide data.
Panel A: Decadal radiocarbon $^{14}$C global production rate \citep{roth13} with the 95\% confidence interval plotted in gray.
Panel B: Quasi-decadal variability of $^{10}$Be flux in the GRIP ice core \citep{yiou97}.
The formal $1\sigma$ error of 7\% (relative to the given value) is indicated by the error bar next to the legend.
}
\label{Fig:dat}
\end{figure}

We used two sets of cosmogenic radionuclide data ($^{14}$C in tree trunks  and $^{10}$Be in polar ice;
 panels A and B in Fig.~\ref{Fig:dat}, respectively) as tracers of solar activity \citep[e.g.,][]{beer12,usoskin_LR_13}.

%


Radiocarbon $^{14}$C is produced in the terrestrial atmosphere by cosmic rays and then takes part in the global carbon cycle
 \citep[e.g.,][]{bard97,beer12,roth13}.
The measured quantity, the relative concentration $\Delta^{14}$C of radiocarbon in tree rings, needs to be corrected
 for the apparent decay and for the carbon cycle effect to reconstruct the $^{14}$C production rate.
Here we used the $^{14}$C production rate, $Q(^{14}$C), as reconstructed by \citet{roth13} for the Holocene, using the
 globally averaged INTCAL09 \citep{reimer_09} radiocarbon database and the dynamical BERN3D-LPJ carbon cycle model, which is
 a new-generation carbon-cycle climate model, featuring a 3D
 dynamic ocean, reactive ocean sediments, and a 2D atmosphere component coupled
 to the Lund-Potsdam-Jena dynamic global vegetation model.
The data were reduced to the decadal temporal resolution.
For the decades around years 775 AD and 994 AD, the production rate was corrected to remove the modeled
 contribution that is due to the occurrence of two extreme solar particle events (\citet{usoskin_775_13};
 see also the discussion in \citet{miyake12}, \citet{miyake13} and \citet{bazilevskaya14}).
Finally, we only consider data before 1900 AD because of the Suess effect, which is related to
 extensive burning of fossil fuel,
 which dilutes radiocarbon in the natural reservoirs and makes the use of the $^{14}$C data after
 1900 more uncertain.

Following \citet{usoskin_AAL_14}, we used 1000 individual realizations of the $Q(^{14}$C) ensemble to describe the consequences
 of uncertainties in the data and carbon cycle modeling \citep{roth13}.
The corresponding production rate is shown in Fig.~\ref{Fig:dat}a (mean (black)
 and 95\% range (gray shaded area) of the 1000 realizations).


The radionuclide $^{10}$Be is produced by cosmic rays in the atmosphere through
 spallation reactions \citep{beer12}.
It becomes attached to aerosols and is relatively quickly precipitated to the ground.
Because of this fast precipitation, it is not completely mixed in the atmosphere and is subject to some complicated transport.
We relied on the parameterization of the atmosphere transport and deposition of
 beryllium proposed by \citet{heikkila09}.
Here we used a long series of $^{10}$Be depositional flux measured in central Greenland in the framework of the Greenland Ice Core Project (GRIP) for the period before 1645 AD \citep{yiou97}.
We considered the mean data set reduced to quasi-decadal time resolution.
The corresponding rate is shown in Fig.~\ref{Fig:dat}B, where we also plot
 a formal $1\sigma$ uncertainty error (estimated to be 7\% in relative terms, \citet{yiou97}).

\subsection{Axial dipole evolution over the past 9000 years}
\label{Sec:geom}
Two approaches can be used to constrain the axial dipole moment evolution over the past few millennia.
The first consists of constructing global geomagnetic field models in the form of time-varying series of Gauss coefficients by
 taking advantage of all available archeo- and paleomagnetic data \citep[see, e.g.,][]{korte05,korte11b,korte11a,licht13,pavon14,nilsson14}
 and using the corresponding axial dipole component $| {g^1_0} |$.
Differences among these models mainly come from the treatment applied to the data, in particular the way
 experimental and dating uncertainties are being handled (see discussion and details in the references above).
In the present study, we considered three recent models, referred to as A\_FM \citep{licht13},
 SHA.DIF.14k \citep{pavon14}, and pfm9k.1a/b \citep{nilsson14}.
We note that A\_FM and SHA.DIF.14k were built using archeomagnetic and volcanic data sets, whereas
 pfm9k.1a/b also took sedimentary data into account.
We assume that pfm9k.1a/b supersedes the slightly older CALS10k.1b field model
 constructed by \citet{korte11a} using practically the same dataset.

The second approach is based on archeo- and paleointensity data collected worldwide, using archeological
 artifacts and volcanic rocks. The data are transformed into virtual axial dipole moments (VADM), which are then
 carefully weighted to produce a worldwide average VADM.
To be valid, this ``paleomagnetic'' approach requires a dual averaging of the data, both in time and space,
 to best smooth out non-dipole field components \citep[for a discussion, see, e.g.,][]{korte05,genevey08}.
Here we consider the two most recent mean VADM curves built in this way, one by
 \citet{genevey08}, which encompasses the past 3000 years,
 and a second one by \citet{knudsen08}, which covers the entire Holocene.
In addition, and because quite a large number of additional intensity data have recently been collected, an updated
 mean VADM curve was also produced for the purpose of the present study.
For this we used the GEOMAGIA50.v3 data base \citep{brown15}, to which we
added or modified about 390 individual intensity data points (see more details in Appendix A).
The new data compilation contains 4764 intensity values dated to between 7000 BC and 2000 AD.

To build this new VADM curve, we first carried out a series of computations to explore the effects of changing the width
 of the temporal averaging sliding windows (200, 500, and 1000 years) and the size of the region of spatial weighting (over regions
 of 10$^\circ$, 20$^\circ$ , and 30$^\circ$ width).
We also used a bootstrap technique to simulate the effect of noise in the intensity data within their age uncertainties
 and within their two standard experimental error bars 2$\sigma$ \citep[see, e.g.,][]{korte09,thebault10}.
For each set of parameters, an ensemble of 1000 individual curves was computed, allowing us to obtain
 at each epoch the mean VADM and its standard deviation, together
 with the maximum and minimum VADM from the 1000 possible values.
Results from different computations are shown in Appendix A.
This analysis revealed that VADMs derived using sliding windows with widths
 of 500 years and 1000 years are very similar.
VADM values also appear to be relatively insensitive to the size of the area chosen for the regional averaging.
Some differences, but still quite limited, are observed when the width of the sliding window is reduced to 200 years,
 which reveals enhanced variations compared to that obtained when using sliding windows of larger widths, as expected.
Averaging over such a narrow window, however, is reasonable only for the most recent time interval
 (here, the past 3500 years),
 which is documented by a rather large number of data points (3756 among the 4764 available data)
 with a relatively wide but still uneven geographical distribution \citep[see for instance Fig. 4 in][]{knudsen08, genevey08}.
Because of this, we finally decided to build a composite VADM variation curve, which we hereafter refer to as GMAG.9k.
This curve was computed for every 10 years, using sliding windows of widths 200 years
 between 1500 BC and 2000 AD and 500 years between 7000 BC and 1500 BC, with a spatial weighting over
 regions of 10$^\circ$ in size for both time intervals (numerical values for this composite curve are
 provided in the CDS, Tables C.1 and C.2).

\begin{figure}[t]
\centering \resizebox{\columnwidth}{!}{\includegraphics{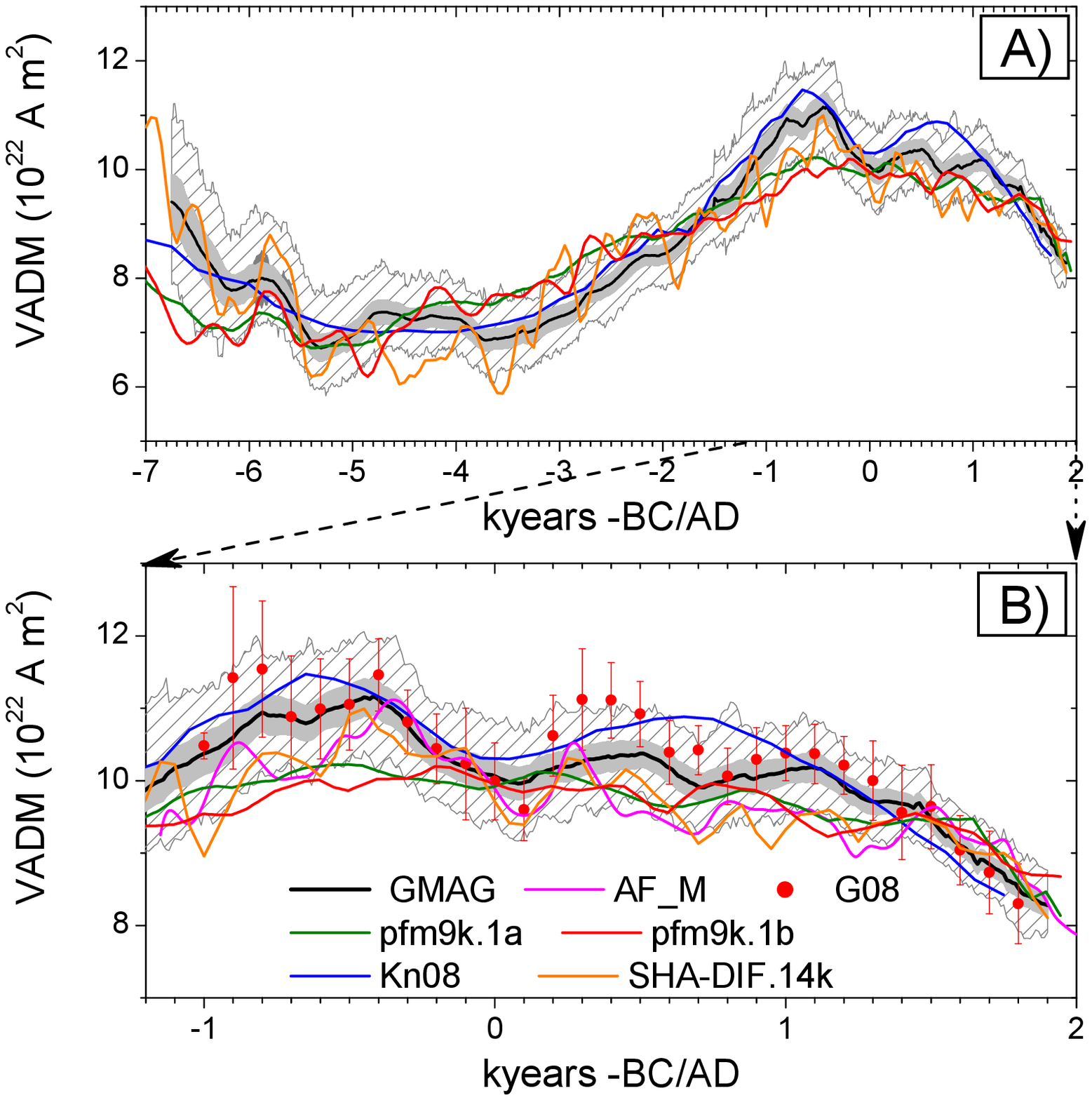}}
\caption{Time series of the axial dipole moment reconstructions spanning the past 9000 years (panel A),
with a zoom for the last 3200 years (panel B).
The black solid line depicts GMAG.9k (the reconstruction presented and used in this work) with $\pm 1\sigma$ and the full range variability
 presented by the gray shading and the hatching, respectively.
Other reconstructions shown are \citep{licht13} denoted as AF\_M, \citep{genevey08} denoted as G08, \citep{nilsson14}
 denoted as pfm9k.1b and pfm9k.1a, \citep{knudsen08} denoted as Kn08, and \citep{pavon14} denoted as SHA-DIF.14k.
For better readability, error bars were omitted for these curves, but this does not affect the discussion of the results (see text).
}
\label{Fig:VADM}
\end{figure}

Figure~\ref{Fig:VADM} shows this new GMAG.9k curve.
This updated VADM curve does not markedly differ from previous dipole moment curves.
Its behavior over the past 3000 years is very similar to that of the VADM curve obtained by \citet{genevey08},
 who also used a spatial weighting, but with a smaller number of different and distant regions
 (30$^{\circ}$ in size) and a smaller dataset selected based on specific quality criteria.
We note, however, that the new VADM curve tends to lie slightly below that of \citet{genevey08}.
Differences with the VADM variation curve of \citet{knudsen08} are larger.
However, the latter was
 computed using sliding windows of width 500 years with no geographical weighting (these authors
 concluded that they had no significant bias despite the poor spatial data distribution).
Comparison with the A\_FM, SHA-DIF.14k and pfm9k.1a/b curves \citep{licht13,pavon14,nilsson14}
 reveals a fairly similar evolution except for a small offset between $\approx 500$ AD and $\approx 1500$ AD.
This offset also persists when a sliding-window duration of 500 years is used for the VADM computations.
However, the A\_FM, SHA-DIF.14k and pfm9k.1a/b curves generally lie within the envelope of possible VADM
 values (Fig.~\ref{Fig:VADM}).
The same observation holds for all the periods before 1000 BC (Fig.~\ref{Fig:VADM}A).
This encouragingly suggests that relying on the GMAG.9k ensemble of 1000 individual VADM curves
 to reconstruct the solar activity as done in the present study, can be considered as a conservative procedure from a geomagnetic point of view.

\section{Reconstructing the solar activity}

Since details of the reconstruction of solar activity from cosmogenic nuclides are described
 elsewhere \citep[e.g.,][]{beer12,usoskin_LR_13}, we only briefly describe this reconstruction and recall important relevant information.
Cosmogenic isotopes are produced by cosmic rays in the terrestrial atmosphere.
Since cosmic rays are modulated by solar magnetic activity, the
variability of cosmogenic isotope production
 reflects the latter.
However, two terrestrial processes may disturb this relation.
One process is additional shielding of Earth from cosmic rays by the geomagnetic field, whose
 changes must be known independently.
For this purpose, we relied on the GMAG.9k axial dipole evolution constructed as described in Sect. \ref{Sec:geom}.
As discussed in Sect.~\ref{S:14C}, another important process is transport and deposition of the nuclides
 in the terrestrial system.
Because of the poorly known details of climate variability in the past, the related transport models
 are commonly adjusted to modern conditions, which may lead to some uncertainties in the older part of the time interval.

We converted the cosmogenic isotope production rate to the Galactic cosmic ray (GCR) flux variability
 using recent production models.
The global production of $^{14}$C was modeled using the model of \citet{kovaltsov12}, while the production of $^{10}$Be
 was modeled using an updated version of the model of \citet{kovaltsov10}.
Cosmic ray variability was calculated in terms of the heliospheric modulation potential
 \citep[see definitions and formalism in][]{usoskin_Phi_05}, also considering  $\alpha$-particles and heavier
 species of cosmic rays \citep{webber09}.
This modulation potential was furthermore converted into decadal (solar-cycle averaged) sunspot number through
 the open solar magnetic flux model \citep{solanki00,krivova07}, which relates the solar surface
 magnetic cycle to the emerging magnetic flux \citep{cameron15}.
We note that the overall reconstruction method used here is similar to that previously
 used by \citet{usoskin_AAL_14}.

Uncertainties were assessed straightforwardly by computing a large ensemble of individual reconstructions.
We used the set of 1000 time-varying individual archeomagnetic reconstructions of GMAG.9k
 (see Sect.~\ref{Sec:geom}), which account in particular for experimental and age uncertainties.
This ensemble was cross-used with a similar ensemble of 1000 production rates of $^{14}$C to account
 for measurement and compilation uncertainties in the IntCal09 and SHCal04 data, in the air-sea gas exchange rate,
 in the terrestrial primary production, and in the closure of the atmospheric CO$_2$ budget \citep{roth13}.
GMAG.9k was also cross-used in the same way with a set of 1000 $^{10}$Be series.
In that case, however, $^{10}$Be series of decadal values
 were generated around the mean provided by GRIP using normally distributed random numbers
 \citep[with a standard deviation equal to 7\% of the mean value,][]{yiou97}) to reflect known errors.
In both cases, all possible combinations of the ensembles yielded $10^6$ series of the reconstructed heliospheric modulation potential,
 next converted into $10^6$ series of sunspot numbers.
These series reflect the error propagation through all the intermediate steps.
An additional random error with $\sigma$=0.5 was finally added to each computed sunspot number to account for the small
 possible error related to the conversion between the modulation potential and the solar open magnetic flux \citep{solanki_Nat_04}.

\begin{figure}[t]
\centering \resizebox{\columnwidth}{!}{\includegraphics{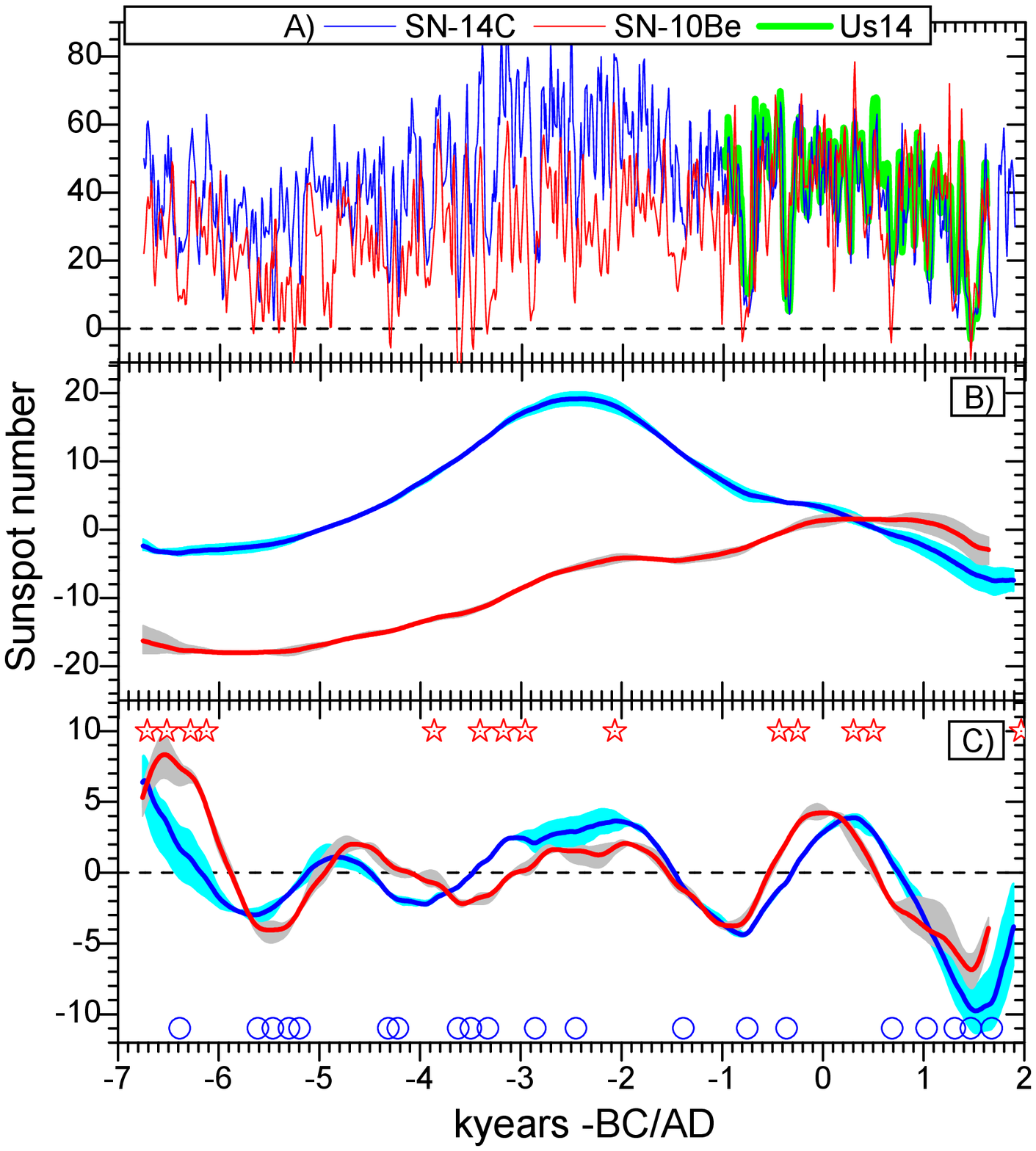}}
\caption{Panel A: Raw reconstructions of the sunspot numbers (mean curves) SN-14C (blue)
 and SN-10Be (red), compared to the recent 3 kyr reconstruction \citep[][ -- green curve]{usoskin_AAL_14}.
Panel B: First component of the singular spectrum analysis (SSA -- see Appendix~\ref{Sec:SSA}) for the SN-14C (blue) and SN-10Be (red) series.
The shaded areas depict the uncertainties related to the parameter $L$ of the SSA.
Panel C: Same as in panel B, but for the second SSA components of the SN-14C (blue) and SN-10Be (red) series.
The large dots and red stars denote times of the grand minima (see Table~\ref{Tab:min}) and
 grand maxima (Table~\ref{Tab:max}), respectively.
}
\label{Fig:SN_c_be}
\end{figure}

Decadal sunspot numbers reconstructed in this way from the $^{14}$C and $^{10}$Be data are henceforth denoted as
 SN-14C and SN-10Be, respectively.
Ensemble means of these SN-14C and SN-10Be are shown in Fig.~\ref{Fig:SN_c_be}A.
These reconstructions agree well with the earlier reconstruction \citep{usoskin_AAL_14}
 that cover the last 3000 years.

\begin{figure}[t]
\centering \resizebox{\columnwidth}{!}{\includegraphics{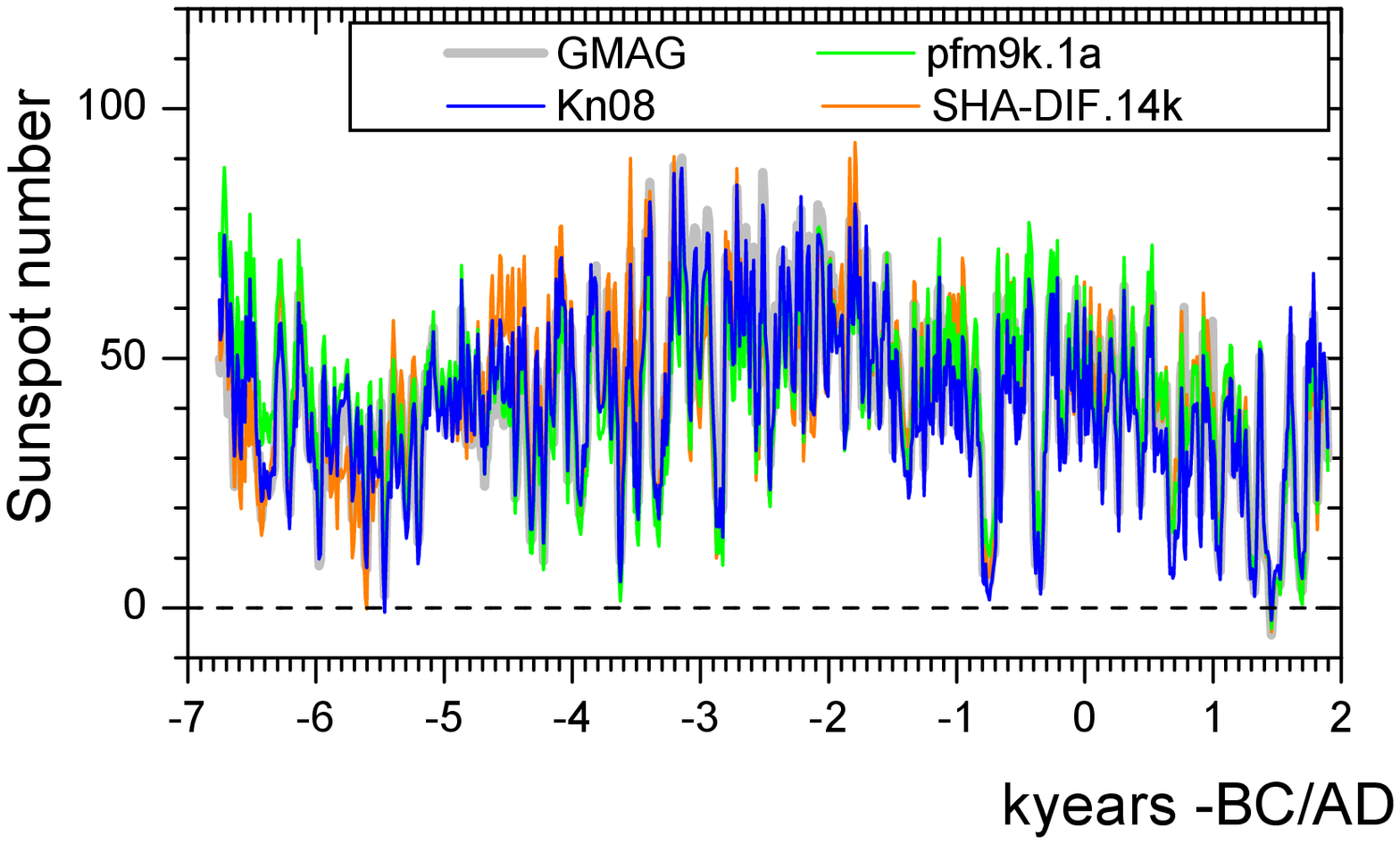}}
\caption{Comparison of alternative SN-14 sunspot number reconstructions when relying on different axial dipole reconstructions,
 (same notations as in Fig.~\ref{Fig:VADM}).
Only the mean curves of the corresponding ensembles are shown.
}
\label{Fig:SN_all}
\end{figure}

We checked the influence of the choice of axial dipole moment reconstruction on the SN-14C
 reconstruction by also considering alternative geomagnetic field models.
Results are shown in Fig.~\ref{Fig:SN_all}.
This figure clearly shows that all $^{14}$C-based SN reconstructions lie close to each other and reveal a common general pattern.
In particular, Figs.~\ref{Fig:SN_c_be}A and~\ref{Fig:SN_all} both display a long-term trend in the $^{14}$C-based SN
 reconstructions over the past 9000 years.
This trend, however, is different from that of the $^{10}$Be-based SN reconstruction.
What causes these long-term trends is unclear: they may reflect various combinations of climate effects
 or improperly corrected geomagnetic field effects.
We discuss this important point in the next section.

\section{Long-term behavior}

To investigate the long-term behavior of the reconstructed solar activity, we relied on the singular
 spectral analysis (SSA) method \citep{vautard89, vautard92} as described in Appendix B.
This SSA was applied to both the SN-14C and SN-10Be series, and the corresponding two first SSA
 components are shown in Fig.~\ref{Fig:SN_c_be}B and C.
The robustness of these SSA results has also been assessed by considering a wide range of values for the embedding
 dimension $L$.
The resulting uncertainties are indicated by means of shaded areas in these figures.

\subsection{Multimillennial trend: Possible climate influence}
\label{Sec:S1}

Here, we first consider the long-term primary SSA components of the SN-14C and SN-10Be series.
These are shown in Fig.~\ref{Fig:SN_c_be}B.
The shaded areas show the full range of computations for $L$-values ranging between 150 and 200
 for $^{14}$C and between 120 and 170 for $^{10}$Be.
These primary components are well identified.

\begin{figure}[t]
\centering \resizebox{\columnwidth}{!}{\includegraphics{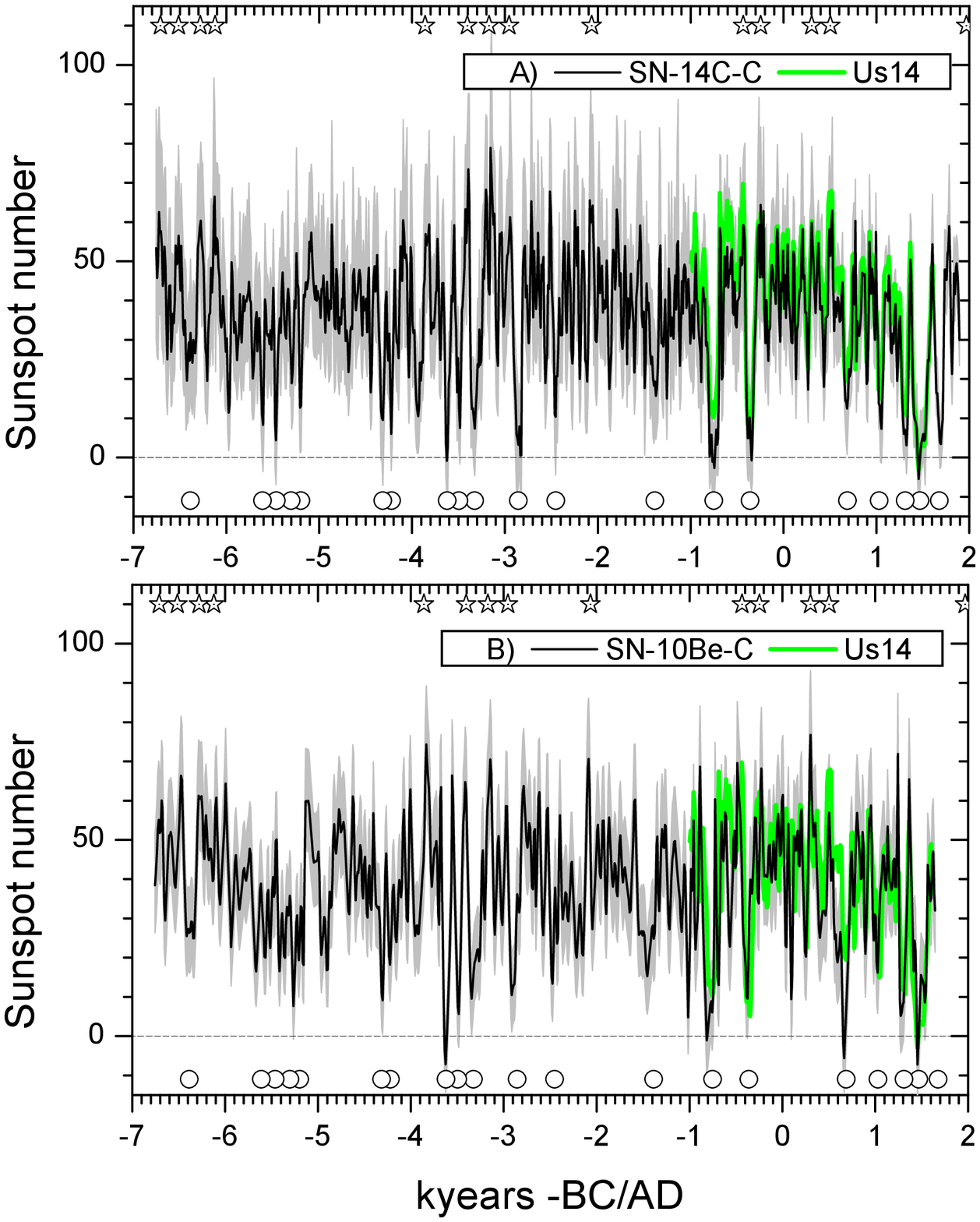}}
\caption{Corrected sunspot number reconstructions SN-14C-C (panel A) and SN-10Be-C (panel B), after removing
 long-term trends (see Sect.~\ref{Sec:S1}).
 The black curves and the gray shading depict the mean and the 95\% range (over $10^6$ ensemble members) of the reconstructions,
  respectively, and the green curve represents the 3 kyr reconstruction by \citet{usoskin_AAL_14}.
  Stars and circles denote grand maxima and minima, respectively, as in Fig.~\ref{Fig:SN_c_be}.
  Tables for this plot are available at the CDS, including the mean values and the uncertainties of the
   sunspot numbers reconstructed here as shown in Fig.~\ref{Fig:SN-C}.
  }
\label{Fig:SN-C}
\end{figure}

Just as clearly, it also appears that these primary components are different for the two series: SN-14C yields a
 single, nearly symmetric wave along the entire time interval of 9 millennia with a range of about 20 in
 sunspot number, while SN-10Be yields a nearly monotonous trend within the same range of about 20 in sunspot number.
The fact that these trends differ so much implies that they can hardly be related
to a common process.
This makes terrestrial processes, in particular transport and deposition,
a much more likely cause.
Differences in the very long term Holocene trends between the two isotopes have previously been noted
 and ascribed to such terrestrial processes  \citep{vonmoos06,usoskin_10Be_09,steinhilber12,inceoglu15}.
Climate change, in particular, is a likely cause because it affects the two isotopes in very different ways
 \citep{beer12}, with $^{14}$C being sensitive to long-term changes in the ocean circulation
  \citep[e.g.,][]{hua15}, while $^{10}$Be is mainly sensitive to large-scale atmospheric dynamics.
In any case, it is quite clear that these long-term trends are unlikely to be of solar origin.
For this reason, we decided to remove them from our original SN-14C and SN-10Be series to produce
 what we hereafter refer to as the SN-14C-C and SN-10Be-C series, where
 the last 'C' stands for 'corrected'.
The corresponding solar activity reconstructions are shown in Fig.~\ref{Fig:SN-C}.
(We note that since these reconstructions were corrected for long-term trends, they only reflect relative changes within the solar activity, strictly speaking).

\begin{figure}[t]
\centering \resizebox{\columnwidth}{!}{\includegraphics{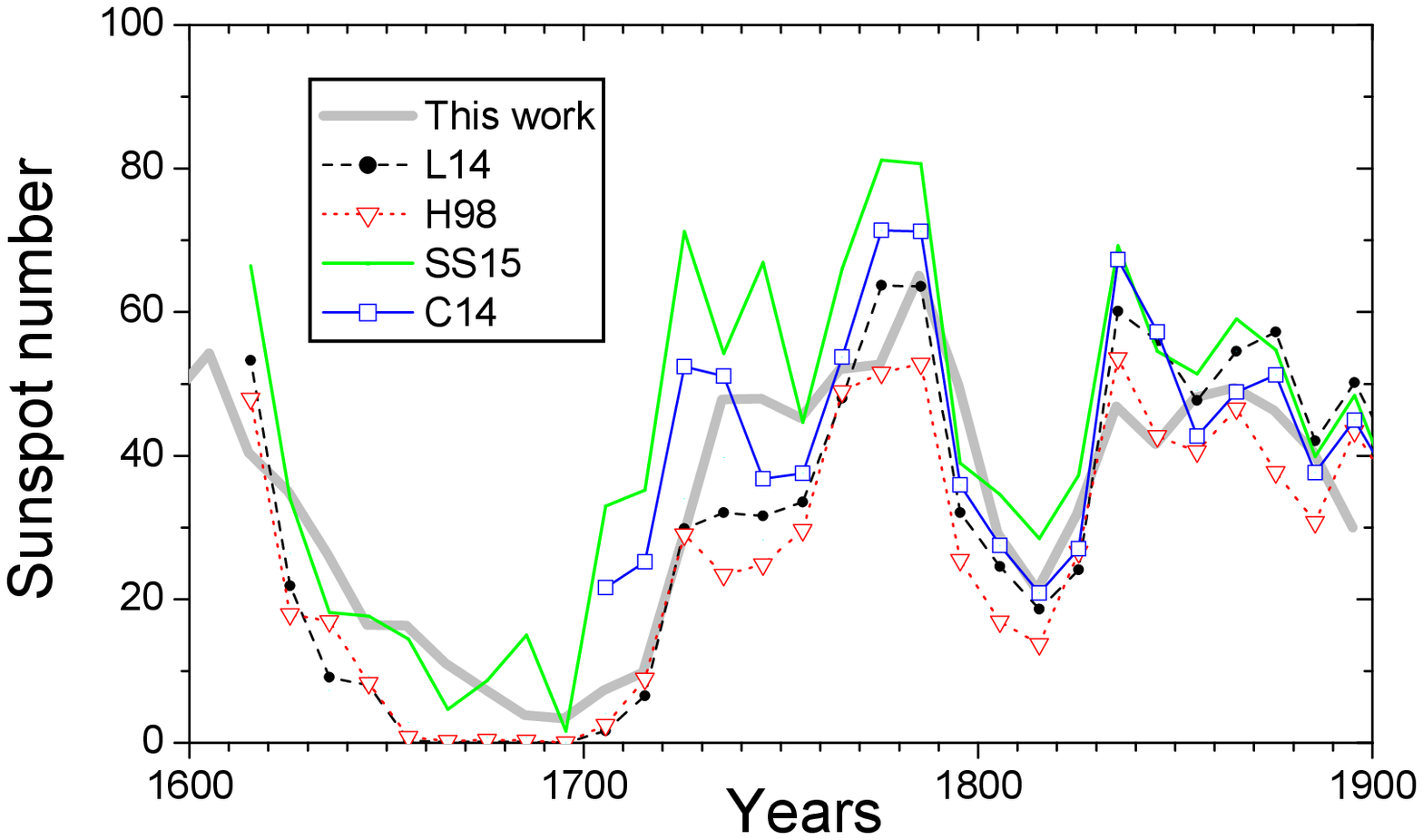}}
\caption{Decadally averaged sunspot numbers for the period 1600--1900 AD.
 The thick gray curve represents this work (uncertainties are not shown).
 Other curves correspond to sunspot number series: L14 \citep{lockwood_1_14},
 H98 \citep{hoyt98}, SS15 \citep{svalgaard15}, and the C14 international
 sunspot number (v.2) scaled with a factor 0.6 \citep{clette14}.
  }
\label{Fig:comp_1600}
\end{figure}
We also show in Fig.~\ref{Fig:comp_1600} the comparison of the sunspot number series
 for the period 1600\,--\,1900 AD reconstructed
 here from $^{14}$C with other series based on sunspot counts and
 drawings \citep{hoyt98,lockwood_1_14,clette14,svalgaard15}.
All series exhibit the same centennial-scale evolution, while at shorter timescales the variations
 differ from one series to the next.
However, the reconstruction clearly lies within the range of the other sunspot series.

\subsection{Common signature of the Hallstatt cycle}

We now consider the second SSA components as shown in Fig.~\ref{Fig:SN_c_be}C.
These components are dominated by a $\approx 2400$ yr periodicity that is remarkably coherent between the SN-14C and SN-10Be series.
The formal Pearson correlation coefficient between the two curves shown in Fig.~\ref{Fig:SN_c_be}C is $0.77\pm0.01,$ which is highly
 significant with $p<10^{-5}$ estimated using the non-parametric random phase method \citep{ebisuzaki97,usoskin_JASTP_06}.

\begin{figure}[t]
\centering \resizebox{\columnwidth}{!}{\includegraphics{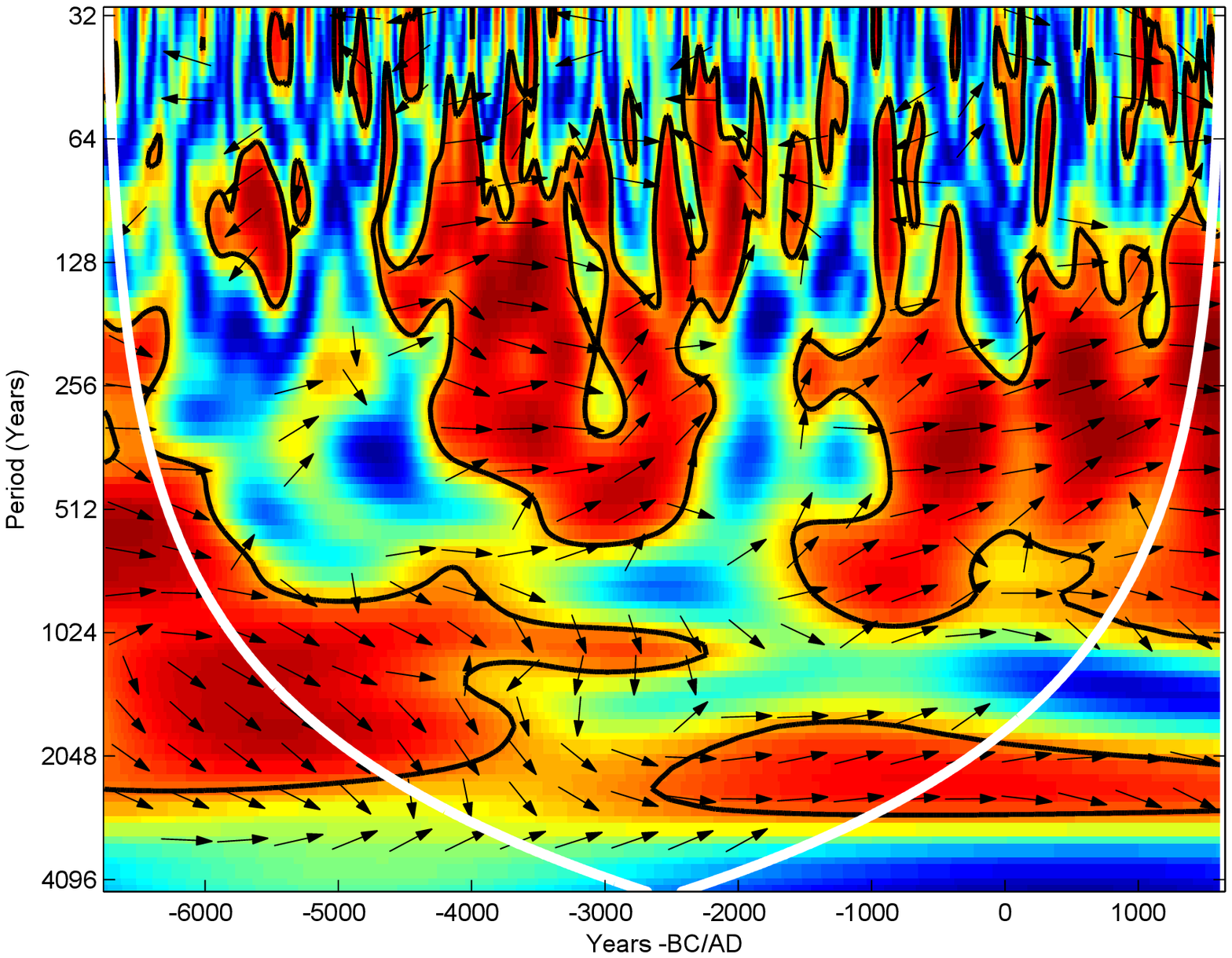}}
\caption{Wavelet coherence between the SN-14C and SN-10Be series.
The color code gives the value of the coherence from 0 (blue) to 1 (red).
The arrows denote the relative phase between the series so that the right-pointing arrows correspond to an exact in-phase
and the left-pointing arrows to an exact anti-phase relation.
Black contours delimit the areas of high coherence (95\% confidence level).
White curves delimit the cone of influence where results can be influenced by the edges of the time series
 (beyond which the analysis is possibly biased).}
\label{Fig:wv}
\end{figure}

To confirm the significance of this observation, we also carried out a wavelet coherence analysis of the SN-14C and SM-10Be series.
The wavelet coherence is a normalized cross-spectrum of the two series and provides a measure of their covariance in
 time-frequency domain. It was calculated using the Morlet basis and a code originally provided by \citet{grinsted04}, but
 modified to adopt the non-parametric random phase method for assessing confidence level \citep{ebisuzaki97,usoskin_JASTP_06}.
The corresponding wavelet coherence is displayed in Fig.~\ref{Fig:wv}.
It shows that while the coherence is strong but intermittent at shorter timescales and nearly absent at the longest timescales
 \citep[cf.][]{usoskin_10Be_09}, there is a wide band of very high coherence that is in phase along the entire
 time interval for the periods of 2000--3000 years, consistent with the periods seen in the second SSA component plotted in
 Fig.~\ref{Fig:SN_c_be}C.

It is therefore clear that the $\approx 2400$ yr quasi-periodicity is common to both series and that it dominates their
 super-millennial timescale variability.
It is related to the so-called Hallstatt cycle that is known in $\Delta^{14}$C \citep[e.g.,][]{damon91,vassiliev02,ma15},   but has been poorly documented until now in the $^{10}$Be data \citep{mccracken11,hanslmeier13}.
We also note that the principal component analysis applied by \citet{steinhilber12} to a composite solar activity reconstruction
 likewise revealed a Hallstatt cycle in the heliospheric modulation potential that is synchronous with the reconstruction
 shown in Fig.~\ref{Fig:SN_c_be}C, although it was not explicitly characterized.

This Hallstatt cycle has so far either been ascribed to climate variability \citep{vassiliev02} or to geomagnetic fluctuations,
 particularly geomagnetic pole migration \citep{vasiliev12}.
However, the fact that the signal we found is in phase and of the same magnitude in the two cosmogenic isotope reconstruction implies that
 it can hardly be of climatic origin.
As already pointed out, $^{14}$C and $^{10}$Be respond differently to climate changes.
In particular, $^{14}$C is mostly affected by the ocean ventilation and mixing, while $^{10}$Be (in particular its
 deposition on central Greenland) is mainly affected by the large-scale atmospheric circulation, particularily in
 the North Atlantic region \citep{field06,heikkila09}.
It can also hardly be of geomagnetic origin and related to geomagnetic pole migration, since  $^{14}$C is
 globally (hemispherically) mixed in the terrestrial system and insensitive to the migration of these poles.
To reproduce the observed Hallstatt cycle, the dipole moment (VADM) would have to vary, with the corresponding period,
 in a range of $\approx 2\times 10^{22}$ A m$^2$ (i.e., by about 20\%).
This is not supported by any geomagnetic field reconstruction (Fig.~\ref{Fig:VADM}).
The SSA analysis of the geomagnetic series (not shown) does not yield the Hallstatt cycle.

We thus conclude that the $\approx 2400$-yr Hallstatt cycle is most likely a property of the long-term solar activity.

\begin{figure}[t]
\centering \resizebox{\columnwidth}{!}{\includegraphics{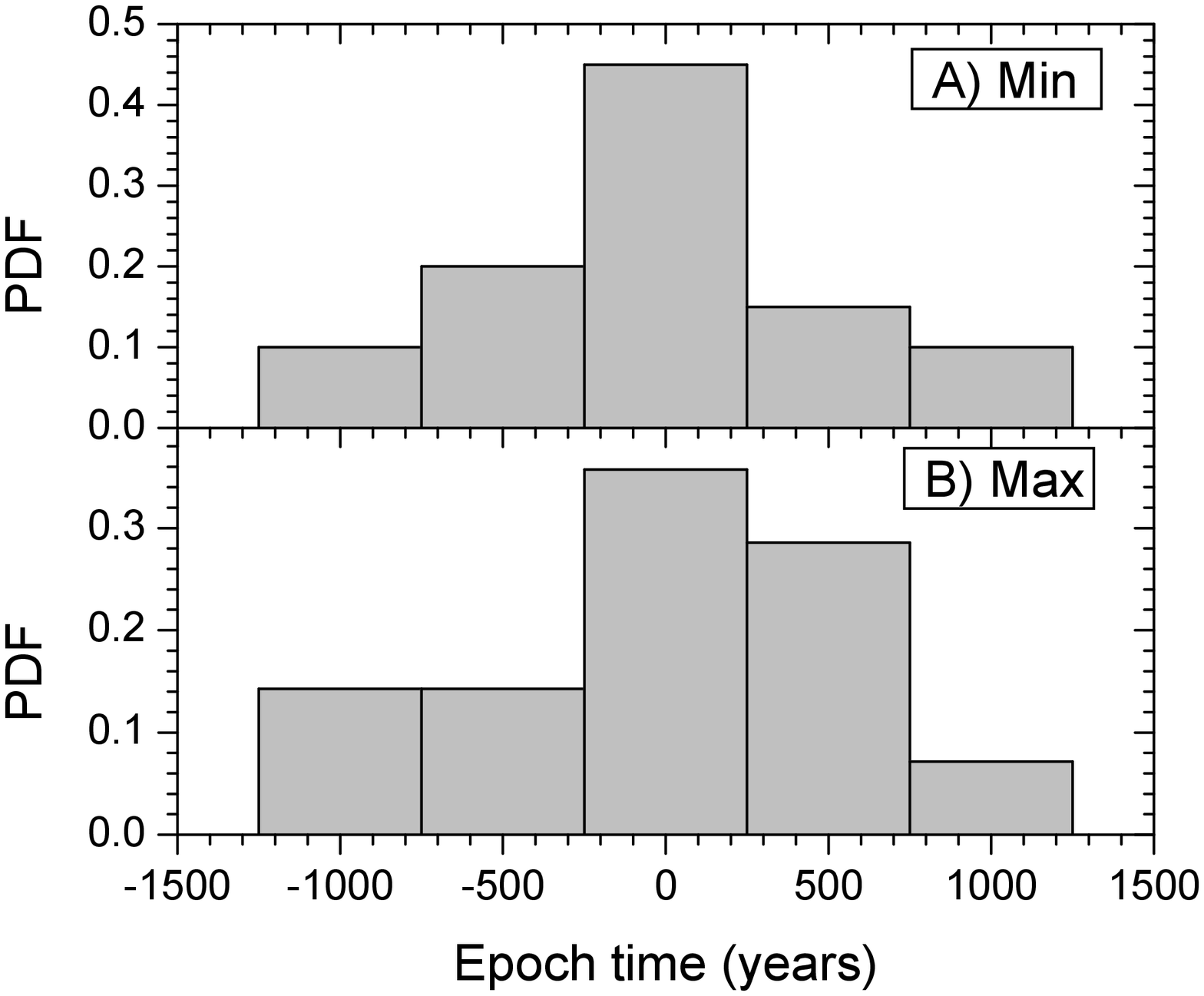}}
\caption{Probability density function (pdf) of the time of occurrence of grand minima (panel A) and
 grand maxima (panel B) relative to the time of occurrence of the nearest high and low, respectively, of the Hallstatt cycle, using the
 superposed epoch analysis.
 The times of occurrence of highs and lows of the Hallstatt cycle are defined by considering the average of the two curves
 (second SSA components of the SN-14C and SN-10Be series)
 shown in Fig.~\ref{Fig:SN_c_be}C
 (leading to  5530 BC, 3650 BC, 810 BC, and 1510 AD for the lows and to 6670 BC, 4600 BC, 1970 BC, and 160 AD for the highs).
 }
\label{Fig:epoch}
\end{figure}


\section{New constraints on the temporal distribution of grand minimum and grand maximum events}

Using the reconstructed SN-14C-C and SN-10Be-C time series shown in Fig.~\ref{Fig:SN-C}, we provide a list of grand minima and
maxima (Tables~\ref{Tab:min} and \ref{Tab:max}, respectively).
In contrast to earlier work \citep{usoskin_AA_07,steinhilber08}, we propose a conservative list based on both the SN-14C-C and SN-10Be-C
 reconstructions, meaning that we only list events that are simultaneously seen in both reconstructions (a time adjustment of the $^{10}$Be-based
 series for $\pm 40$ years was allowed owing to the dating uncertainties \citep{muscheler14}).
To identify grand minima, the following criterion was used (with one exception, see below): the event in both reconstructions
(using the mean of the ensemble) must
 correspond to a SN value below a threshold value of SN=20 for at least 30 years.
Although the event ca. 6385 BC lies slightly above this threshold, we also considered it as a grand minimum
 because it clearly has a Sp\"orer-type (i.e., prolonged) shape and occurs in both series.
It is possible that the level of activity was slightly overestimated for this event because of
 uncertainties in dipole moment evolution during the older part of the time interval.
To identify grand maxima, we similarly requested the events to have a SN value exceeding the threshold of SN=55 for
 at least 30 years in both reconstructions.
We thus defined 20 grand minima with a total duration of 1460 years ($\approx 17$\% of time) and 14 grand
 maxima with a total duration of 750 years ($\approx$8\%).
 These numbers are similar to those estimated earlier by \citet{usoskin_AA_07}, although we note that more grand maxima are now identified.
 In contrast, these numbers are significantly lower than those
recently estimated by \citet{inceoglu15}, who relied on a different type of analysis and set of criteria (somewhat less restrictive,
 leading to the identification of a larger set of events, which is essentially inclusive of the set of events we identified here,
 see Tables~\ref{Tab:min} and \ref{Tab:max}).

Times of grand minima and maxima identified in this way are shown in Figs.~\ref{Fig:SN_c_be}C and \ref{Fig:SN-C}
 as circles and stars, respectively.
An important observation from this figure is that the occurrence of grand minima and maxima appears intermittent in time.
Grand minima seem to be closely related to the Hallstatt cycle, being clearly more numerous during
 the lows of the Hallstatt cycle (see Fig.~\ref{Fig:SN_c_be}C).
This feature was only hinted at in passing by \citet{steinhilber12}.
Figure~\ref{Fig:epoch} shows the pdf (built using the superposed epoch analysis)
 of the grand minima and maxima times of occurrence relative to the time of occurrence of the nearest Hallstatt cycle low and high.
A tendency to cluster is clearly observed.
We checked that a similar tendency also appears when considering the lists of grand maxima and minima provided by
\citet{usoskin_AA_07} and \citet{inceoglu15} over the same time period.
We speculate that this clustering might mean that the probability of a switch of the solar dynamo from
 the normal mode to the grand minimum mode (resp. grand maximum mode), according to \citet{usoskin_AAL_14},
  is modulated by the Hallstatt cycle.
Although this interpretation relies on (necessarily) arbitrary choices for defining grand minimum
 and maximum events, it clearly deserves more investigation to better constrain the behavior of
the solar dynamo on centennial and millennial timescales.


\begin{table}
\caption{List of grand minima with their centers, approximate duration, and comments
( 1 -- listed in \citet{usoskin_AA_07}; 2 -- listed in \citet{inceoglu15}).}
\begin{tabular}{ccl}
\hline
Center & Duration & Comment \\
(-BC/AD) & (years) & \\
\hline
1680 & 80 & Maunder$^\dagger$  \\
1470 & 160 & Sp\"orer \\
1310 & 80 & Wolf  \\
1030 & 80 & Oort \\
690 & 80 & 1, 2 \\
-360 & 80 & 1, 2 \\
-750 & 120 & 1, 2 \\
-1385 & 70 & 1, 2 \\
-2450 & 40 & 2 \\
-2855 & 90 & 1, 2 \\
-3325 & 90 & 1, 2 \\
-3495 & 50 & 1, 2 \\
-3620 & 50 & 1, 2 \\
-4220 & 30 & 1, 2 \\
-4315 & 50 & 1, 2 \\
-5195 & 50 &  2 \\
-5300 & 50 & 1, 2 \\
-5460 & 40 & 1, 2 \\
-5610 & 40 & 1, 2 \\
-6385 & 130 & 1, 2 \\
\hline
\end{tabular}
\\$^\dagger$ independently known.\\
\label{Tab:min}
\end{table}

\begin{table}
\caption{List of grand maxima with their centers, approximate duration, and comments
( 1 -- listed in \citet{usoskin_AA_07}; 2 -- listed in \citet{inceoglu15}).}
\begin{tabular}{ccl}
\hline
Center & Duration & Comment \\
(-BC/AD) & (years) & \\
\hline
1970 & 80 & Modern$^\dagger$ \\
505 & 50 & 2 \\
305 & 30 & 2 \\
-245 & 70 & 2 \\
-435 & 50 & 1, 2 \\
-2065 & 50 & 1, 2 \\
-2955 & 30 & 2 \\
-3170 & 100 & 1, 2 \\
-3405 & 50 & 2 \\
-3860 & 50 & 1,2 \\
-6120 & 40 & 1, 2 \\
-6280 & 40 & 2 \\
-6515 & 70 & 1 \\
-6710 & 40 & 1 \\
\hline
\end{tabular}
\\$^\dagger$ independently known.\\
\label{Tab:max}
\end{table}

\section{Conclusions}

Here we presented new reconstructions of solar activity (quantified in terms of sunspot numbers) spanning the
 past 9000 years (tables are available at the CDS) and assessed their accuracy using different geomagnetic field reconstructions and current
 cosmogenic isotope production models.

We found that the primary SSA components of the reconstructions
that are based on $^{14}$C and $^{10}$Be are significantly different.
These primary components probably reflect long-term changes in the terrestrial system that affect the $^{14}$C and $^{10}$Be isotopes in different ways (ocean
 circulation and/or large-scale atmospheric transport). These components were therefore removed to produce meaningful corrected sunspot number series.
In contrast, the secondary SSA components of the $^{14}$C and $^{10}$Be based reconstructions revealed a common remarkably synchronous $\approx2400$-year quasi-periodicity.
We therefore concluded that this so-called Hallstatt periodicity most likely reflects some periodicity in the solar activity.

From the two cosmogenic isotope records, we finally defined a conservative list of grand minima and grand maxima covering
 the past 9 millennia.
An important finding is that the grand minima and maxima occurred intermittently over the studied period,
 with clustering near maxima and minima of the Hallstatt cycle, respectively.
The Hallstatt cycle thus appears to be a long-term feature of solar activity that needs to be taken into
 account in models of solar dynamo.
 
\acknowledgement{
We are thankful to Raphael Roth and Fortunate Joos for providing data on $^{14}$C production rates and
 for useful discussions on the carbon cycle.
Support by the Academy of Finland to the ReSoLVE Center of Excellence (project no. 272157) and by IPGP
 through its invitation program is acknowledged.
Y.G. and G.K. were partly financed by grant N 14.Z50.31.0017 of the Russian Ministry of Science and Education.
This is IPGP contribution No. 3699. }

\begin{appendix}

\section{GMAG.9k axial dipole evolution}
\label{Sec:App_GMAG}

Here we used the GEOMAGIA50.v3 database  \citep{donadini06,korhonen08,brown15} to which we added
 recent archeo- and paleointensity results \citep{cai14,cai15,cromwell15,degroot15,dichiara14,gallet08,gallet09,gallet10,hong13,kapper15,osete15,shaar15,stillinger15}.
Concerning the data compiled in this version of GEOMAGIA, the Mesopotamian data from \citet{nachasova95,nachasova98}
 were modified according to  \citet{gallet15}.
Further revisions were also discussed in \citet{genevey13} (also A. Genevey, pers. comm.)
 and \citet{gallet15b}.

\begin{figure}[t]
\centering \resizebox{\columnwidth}{!}{\includegraphics{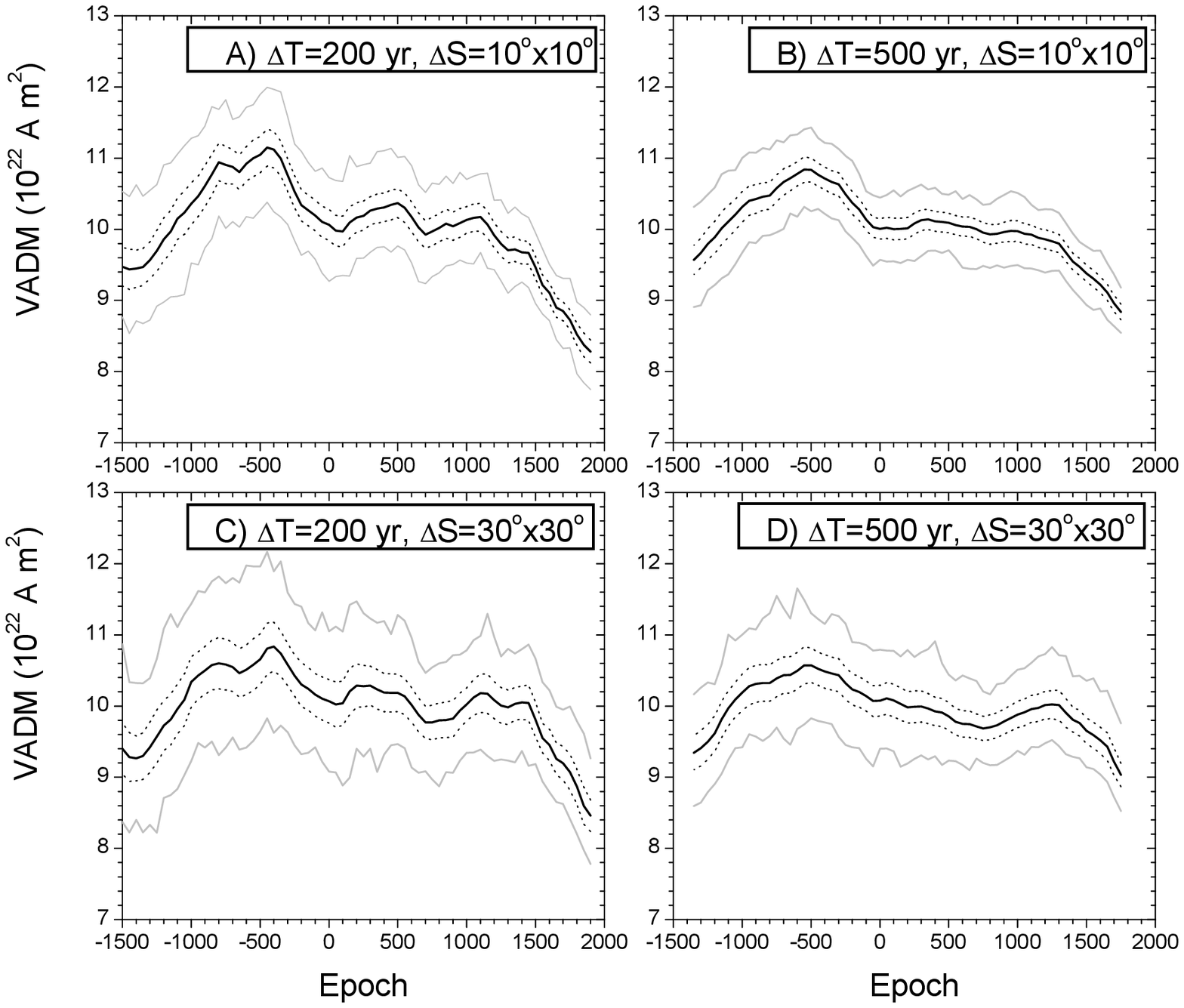}}
\caption{Comparison between VADM curves computed between 1500 BC and 2000 AD using sliding
 windows of $\Delta$T=200 years (panels A, C) and 500 years (B, D) shifted by 50 years, and using a
 weighting over regions of $\Delta$S=$10^\circ\times 10^\circ$ (A, B) and $\Delta$S=$30^\circ\times 30^\circ$
 (C, D) width.
The thick black line exhibits the averaged VADM computed using a bootstrap scheme (see main text and legend
 of Tables C.1 and C.2 available at CDS), with its $1\sigma$ uncertainties (dotted lines) and the envelope of possible VADM values
 (gray lines).
}
\label{Fig:S1}
\end{figure}

\begin{figure}[t]
\centering \resizebox{\columnwidth}{!}{\includegraphics{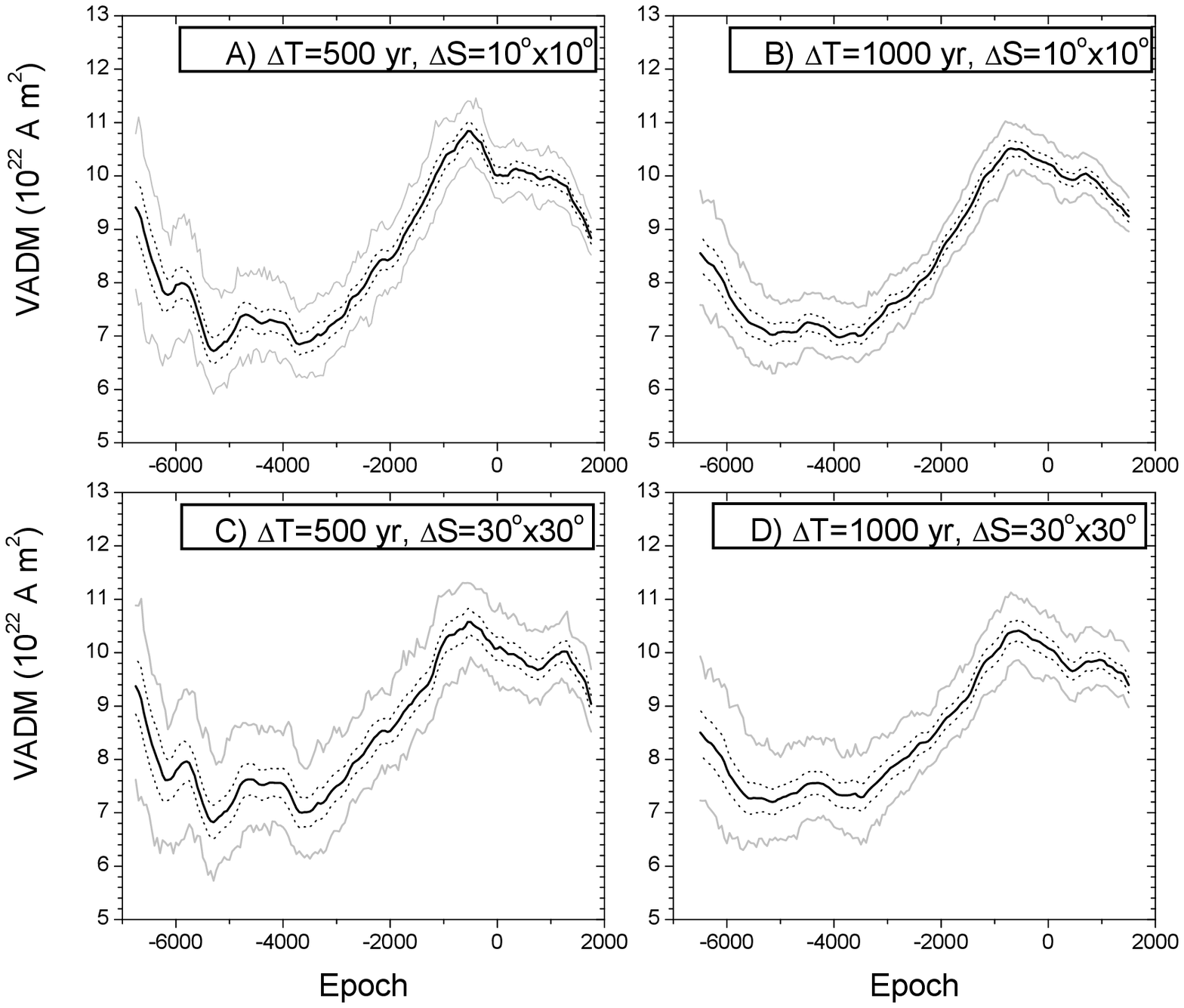}}
\caption{Comparison between VADM curves computed between 7000 BC and 2000 AD using sliding
 windows of $\Delta$T=500 years (panels A, C) and 1000 years (B, D) shifted by 50 years, and using a
 weighting over regions of $\Delta$S=$10^\circ\times 10^\circ$ (A, B) and $\Delta$S=$30^\circ\times 30^\circ$
 (C, D) width.
The thick black line exhibits the averaged VADM computed using a bootstrap scheme (see main text and legend
 of Tables C.1 and C.2 available at CDS), with its $1\sigma$ uncertainties (dotted lines) and the envelope of possible VADM values
 (gray lines).
}
\label{Fig:S2}
\end{figure}

Axial dipole evolution over the past 9000 years, referred to as GMAG.9k, was constrained by
 virtual axial dipole moments (VADM) averaged over sliding windows of 200 years between 1500 BC and
 2000 AD and of 500 years between 7000 BC and 1500 BC shifted by 10 years and using a regional weighting
 scheme over regions of 10$^\circ$ width.
Weights of one third or two thirds were assigned to the regions that contained at a given time interval only one or two
 individual intensity (VADM) data points, respectively.
For those intensity data with no age uncertainties provided in the GEOMAGIA50.v3 database, we used
 the same approach as in \citet{licht13}.
We computed the means of the known age uncertainties over 500 yr long time intervals between
 1000 BC and 2000 AD, over 2000 yr interval between 3000 BC and 1000 BC and over 4000 yr interval
 between 7000 BC and 3000 BC.
After multiplication by a factor of 1.5, the corresponding values were assigned to the intensity data with
 unknown age uncertainties within the periods of concern.
Following \citet{knudsen08}, when no experimental errors were provided on the data, we assigned errors
 amounting 25\% of the corresponding intensity values.
We also note that for the archeomagnetic data for which no attempt was made to take the
 cooling rate effect on thermoremanent magnetization acquisition
into account, we systematically implemented a
 cooling rate correction of 5\% decrease \citep[see for instance in][]{genevey08}.
Mean VADM estimates were then derived using a bootstrap technique to account for the noise in the available
 paleo- and archeointensity data within their age uncertainties and within their $2\sigma$ experimental uncertainties.
1000 VADM curves, also derived using different randomly attributed locations of the weighting regions,
 were hence determined, whose statistics are summarized in Tables C.1 and C.2 (available in CDS) for the periods
 1500 BC-2000 AD and 7000 BC-1500 BC, respectively.
These tables provide for each epoch (first column) a mean VADM (second column), a standard deviation
 (third column), and the maximum and minimum values defining the envelope of possible VADM results
 (fourth and fifth columns).
The variability of VADM is shown in  Figs.~\ref{Fig:S1} (between 1500 BC and 2000 AD) and
 \ref{Fig:S2} (between 7000 BC and 2000 AD).


\section{ Identification of long-term trends by singular spectrum analysis}
\label{Sec:SSA}

The non-parametric singular spectrum analysis (SSA) of time series
 is based on the Karhunen-Loeve spectral decomposition theorem \citep{kittler73} and the Man\'e-Takens
 embedded theorem \citep{mane81,takens81}.
It allows a time series to be decomposed into several components with distinct temporal behaviors and is very
 convenient to identify long-term trends and quasi-periodic oscillations.

The basic version of SSA that we used consists of four straightforward steps
\citep[see, e.g.,][]{Golyandina2001, Hassani2007}:
embedding, singular value decomposition, grouping, and reconstructions.

When considering a real-value time series ${\bf x}$ ($x_1$, $x_2$, $\dots$, $x_{\rm N}$),
the first step of this SSA consists of embedding this series into an $L$-dimensional vector space, using lagged copies of
 ${\bf x}$ to form the so-called trajectory (Hankel) matrix (where $K=N-L+1$ and $L$ is a parameter to be chosen),
\begin{equation}
{\bf X} = \left|
\begin{array}{cccc}
  x_1 & x_2 & \dots & x_{\rm K} \\
  x_2 & x_3 & \dots & x_{\rm K+1} \\
  \dots & \dots & \dots & \dots \\
  x_{\rm L} & x_{\rm L+1} & \dots & \ x_{\rm N}\\
 \end{array}
 \right|
\label{Eq:X}
.\end{equation}
The second step consists of performing a singular value decomposition \citep{golub65} of the trajectory
 matrix. This provides a set of $L$ eigenvalues $\lambda_i$ (arranged in
 decreasing order $\lambda_1\ge\lambda_2\ge\dots\ge\lambda_{\rm L}\ge 0$) and eigenvectors $U_i$
 (often called ``empirical orthogonal functions'') of the matrix  ${\bf D} = {\bf X\, X}^{\rm T}$.
If we denote with $d$  the number of nonzero eigenvalues, we may next define $V_i = {\bf X}^{\rm T}U_i/\sqrt{\lambda}_i$ ($i=1,\dots, d$).
Then, the trajectory matrix can be written as a sum of elementary matrices ${\bf X} = {\bf X}_1+\dots+{\bf X}_{d}$,
 where ${\bf X}_i = \sqrt{\lambda_i}\, U_i\, V_i^{\rm T}$.

Once this decomposition has been completed, the third step consists of the construction of groups of components
by rearranging ${\bf X}$ into ${\bf X} = {\bf X}_{G_1}+{\bf X}_{G_2}+\dots$, where each ${\bf X}_{G}$ is the sum
 (group) of a number of ${\bf X}_i$.
The choice of the components to be considered in each group is made empirically by grouping eigentriples
 ($\sqrt{\lambda},\, U,\, V$) with similar eigenvalues.
Finally, a diagonal averaging is applied to each ${\bf X}_{G}$ to make it take the form of a trajectory matrix, from
 which the associated time series component $\tilde{x}_{G}$ of length N can be recovered
 \citep[for details, see, e.g.,][]{Golyandina2001, Hassani2007}.

\end{appendix}


\end{document}